\documentclass[12pt]{iopart}

\usepackage{amssymb,graphicx}
\usepackage{epstopdf}
\usepackage{dcolumn}
\usepackage{bm}

\begin{document}

\title{Graphene fish-scale array as  controllable reflecting photonic structure}

\author{Victor  Dmitriev$^1$, Clerison Nascimento$^1$ and Sergey L. Prosvirnin$^{2,3}$}
\address{$^1$ Electrical Engineering Department, Federal University of Para, Belem, Para, Brazil.}
\address{$^2$ Institute of Radio Astronomy of National Academy of Sciences of Ukraine, Kharkiv 61002, Ukraine.}
\address{$^3$ Karazin Kharkiv National University, Kharkiv 61077, Ukraine.}
\ead{clerisson@ufpa.br}
\vspace{10pt}

\begin{indented}
\item[]February 2014
\end{indented}

%
%
%

\begin{abstract}
We report resonant features of novel controllable reflectarray which  consists of meander-like graphene strips placed on a metal-backed dielectric substrate. The structure manifests two kinds of resonances appeared as sharp deeps of reflectivity. The first one exists because the strips of periodic cells  of the structure have resonant sizes for induced surface plasmon-polaritons. The second kind of resonances is defined by excitation of TM eigenwaves of the whole structure as a plane photonic crystal. The latter resonances do not depend on whether the strips of the unit cells have resonant sizes  or not.
\end{abstract}

\maketitle
\ioptwocol 

\section{Introduction}

Planar  arrays of conducting elements find many applications, especially as frequency selective surfaces (FSS) and slow-wave structures. In recent years, periodical structures with different types of dielectric and conducting elements became very popular in the design of the so-called electromagnetic crystals also known as photonic band gap structures \cite{Joannopoulos-2008}. 

Two-periodic planar strip arrays are promising for applications because they can possess resonant properties in frequency region of a single-wave regime due to complex shape of their elements. Besides, they have very small thickness. Transmission and reflection properties of the complex-shaped metal strip arrays were studied recently with reference to potential applications of them as filters, thin-film sensors and slow-light devices (see, for example, a review paper \cite{ibraheem-2013-THz_review}). In many of these works, it was supposed that the strip particles placed in different cells of a periodic structure have no electric connection.

The purpose of this work is to study {\em resonant} reflection properties of 2D (two-dimensional) periodic array consisting of long continues {\em graphene} wavy-shaped strips \cite{prosvirnin-2002-ewd} (``fish-scale'' \cite{fedotov-2005-pem}), placed on a metal-backed dielectric substrate. A unit cell of the array is shown in Fig.~\ref{fig:Structure}a. We report resonant features of this array as a planar photonic structure which were not investigated before.

The fish-scale morphology, also called meander-line structure possesses resonant properties. Such metallic gratings are commonly used in optical magnetic mirrors and polarization transformers \cite{prosvirnin-2006-APL, han-2011-FS_polar_transf}. They can serve also as  filters \cite{zhang-2012-BilayerFS}. There are ideas to exploit electromagnetically induced transparency in such structures \cite{papasimakis-2008-mao}. 

Graphene is a 2D material with only one atom thickness which manifests unique electronic and electromagnetic properties  \cite{geim-2007-graphene, grigorenko-2012-graphene}. Electrically tunable carrier concentration of graphene is used in hybrid devices, where  metal plasmonic structures and graphene are combined to broadening and tuning  resonances \cite{ozbay-2013-APL}. Besides, plasmonic resonances directly in the graphene elements provide a great potential for designing devices with a remarkably high absorption \cite{koppens-2011-GraphenePlasmonics, chen-2015-GraphenePlasmonics} and unprecedented levels of field confinement \cite{papasimakis-2013-graphene}. Several pure graphene planar arrays with  different functionality have been suggested recently \cite{Dmitriev:filter,Dmitriev:switch}.

When the graphene structure is placed on a metallic mirror coating with a thin dielectric layer as a substrate it becomes a good broadband reflector everywhere apart from few frequency regions where the reflectivity is small. At these frequencies the reflected wave shows no or a small phase change with respect to the incident wave, thus resembling a reflection from a hypothetical magnetic wall. We also observed that the structure acts as a local field concentrator and a resonant amplifier of losses in graphene strips and underlying dielectric layer. Control bias potential in the fish-scale graphene structure discussed below can be applied via  uninterrupted strips. The bias voltage changes  chemical potential of graphene and consequently, its conductivity. 

We will show that there are two kinds of resonances in  fish-scale graphene reflectarray which differ by physical nature. The first one exists due to strips of the  periodic cell having resonant dimensions for excited surface plasmon polaritons (SPPs). In the case of metal-made fish-scale array, these kind of current resonances is  known and well investigated \cite{fedotov-2005-pem}. 

The second kind of resonances of this  array results in excitation of eigenmodes of the whole structure as a planar photonic crystal. These resonances almost do not depend on whether the graphene strips of the unit cells  have resonant sizes or not. To the best of our knowledge, the resonances of this kind in fish-scale array were not investigated before. 

\section{Description of fish-scale reflectarray}
\begin{figure} 
\centerline{\includegraphics[width=.95\columnwidth]{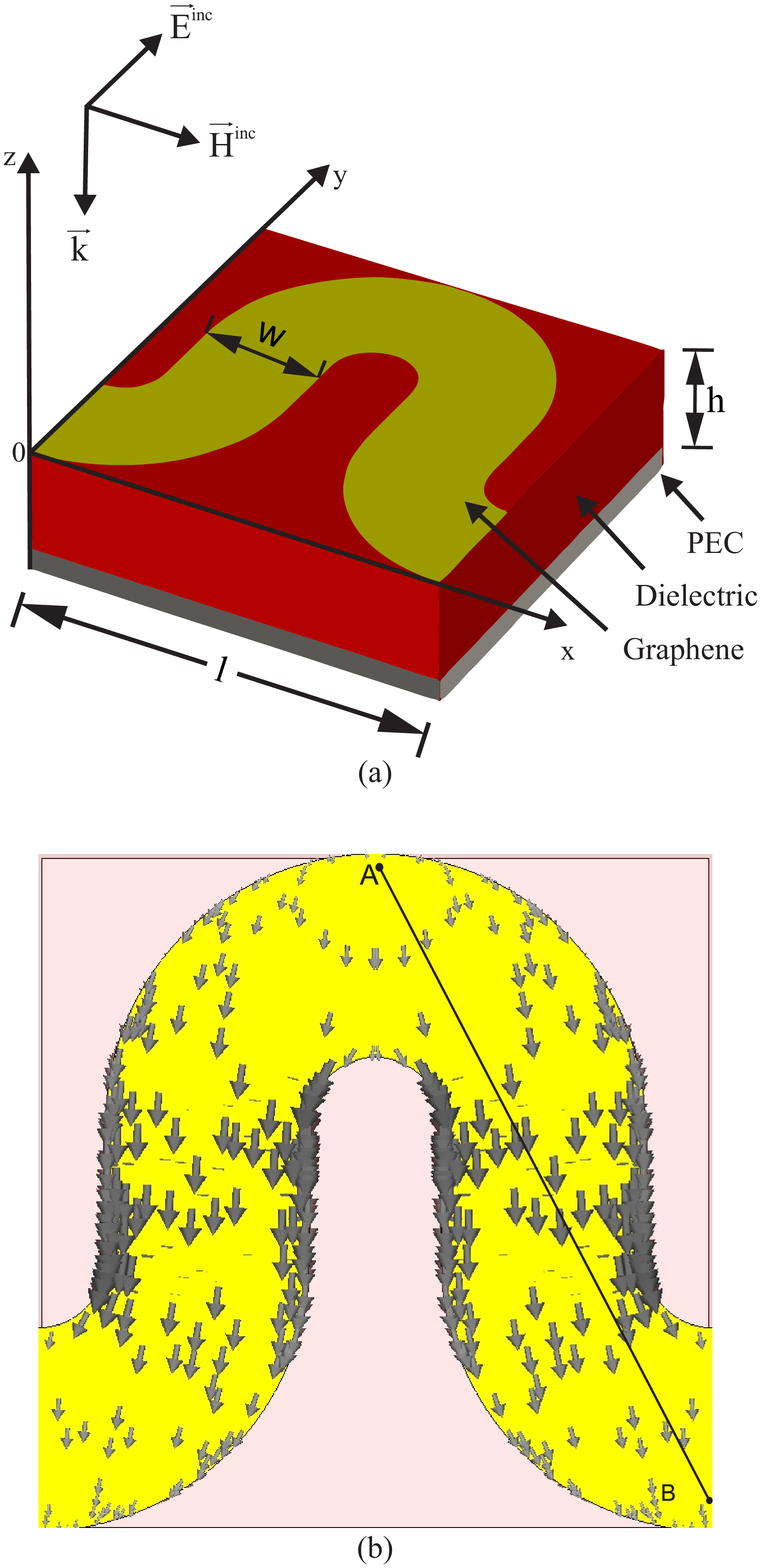}}
\caption{(a) Unit cell of fish-scale graphene array placed on PEC-backed dielectric substrate,  coordinate frame, and main sizes: $l=100~\mu$m, $w=30$~$\mu$m, $h=25$~$\mu$m, $\mu_c=0.5$~eV, resonant frequency is 0.45~THz. (b) Surface current distribution  in  case of excitation of resonance of dimensional kind by normally incident $y$-polarized wave.} \label{fig:Structure}
\end{figure}

Fish-scale graphene array is patterned on the interface of dielectric substrate which has the thickness $h$ and  non-dispersive relative permittivity $\varepsilon=3.5$. A unit cell of the array is shown in Fig.~\ref{fig:Structure}a. The graphene strip has the width $w$, its thickness is infinitely small in comparison with all structure sizes and wavelength. For the sake of simplicity, a square shape of unite cell is chosen. The pitch of the array is $l=100\mu m$. A perfect electric conductor (PEC) screen is placed on the opposite side of the substrate. $\vec{E}^{inc}$, $\vec{H}^{inc}$ and $\vec{k}$  are electric field, magnetic field and wave vector of the incident plane wave, respectively.

\section{SPPs supported by graphene placed on  PEC-backed dielectric substrate}
To study and explain the physical nature of resonant features of graphene reflecting fish-scale array, we start with a discussion of  basic properties of plasmon-polaritons propagating in investigated layered structure without patterning of graphene sheet. 

\begin{figure} 
\centerline{\includegraphics[width=.95\columnwidth]{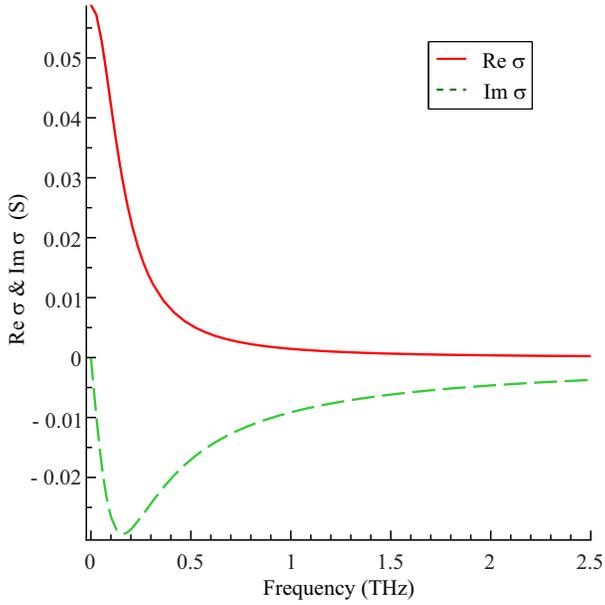}}
\caption{Surface conductivity of graphene Re$\sigma$ (red solid line) and Im$\sigma$ (green dashed line) versus frequency,  $\mu_c=0.5$~eV.} \label{fig:sigma}
\end{figure}

In problems of macroscopic electromagnetics in low THz frequency region, the electric properties of  graphene can be described by  surface conductivity \cite{hanson-2008-JApplPhys}
\begin{equation}
\sigma=-i\frac{e^2k_BT}{\pi \hbar^2(\omega-i2\Gamma)} \left[\frac{\mu_c}{k_BT}+ 2\ln\left(e^{-\mu_c /(k_BT)}+1\right)\right]
\end{equation} 
where $\mu_c$ is the chemical potential, $T$ is the temperature (the value of 300~K was chosen in the simulations), $\Gamma=(2\tau)^{-1}$ is the scattering rate, $\tau$ is the relaxation time (set as 1~ps). Frequency dependence of graphene conductivity is shown in Fig.~\ref{fig:sigma}.

Let us consider  TM polarized SPP supported by graphene. A basic structure is illustrated in the insert of  Fig.~\ref{fig:SPPim}. According to the classical waveguide theory, magnetic field of the SPP propagated along $x$-axis in different regions of the layered structure can be written as follows:
\begin{equation}
\begin{array}{lr}
H_y^1=A e^{-i\beta x -i\gamma_1 z}, & z>0 \\
H_y^2=e^{-i\beta x}[B e^{i\gamma_2 z}+C e^{-i\gamma_2(z+h)}], &  -h<z<0,  
\end{array}
\end{equation}
where $\gamma_1=\sqrt{k^2-\beta^2}$ and $\gamma_2=\sqrt{k^2\varepsilon-\beta^2}$ while Im$\gamma_1<0$ and Im$\gamma_2<0$ with $\beta$ being propagation constant of the SPP, $k$ is the wave number in free space. By using the boundary conditions, that is, the transverse to $z$-axis component of electric field is zero in the plane $z=-h$ and continuous in the plane $z=0$ but the transverse to $z$-axis magnetic field has a discontinuity in this plane which is equal to the value of surface current density on the graphene sheet:
\begin{equation}
H_y^1-H_y^2=-\sigma E_x, \quad z=0,
\end{equation}
the dispersion equation of SPP can be derived:
\begin{equation}
\gamma_1 \varepsilon (1+e^{-2i\gamma_2 h})+ \gamma_2 (1+\sigma \eta \frac{\gamma_1}{k})(1-e^{-2i\gamma_2 h})=0, \label{eqMS}
\end{equation}
where $\eta$ is the impedance of free space.

\begin{figure} 
\centerline{\includegraphics[width=.95\columnwidth]{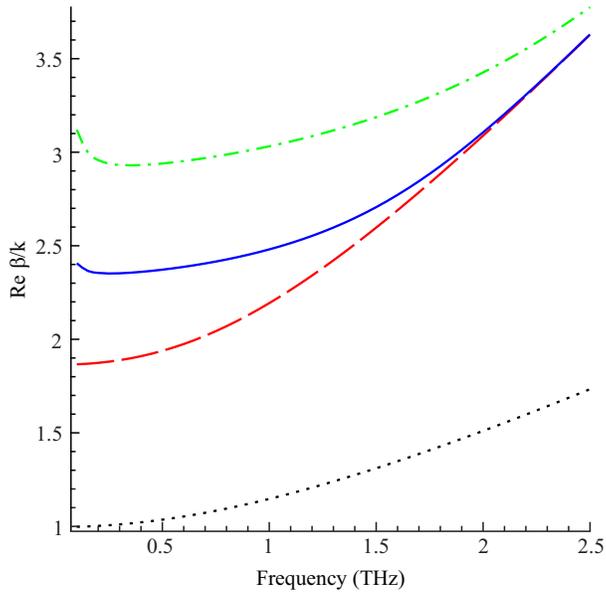}}
\caption{Frequency dependences of real parts of propagation constants of SPPs in  cases of free standing graphene sheet (dotted line), graphene placed on  interface of dielectric semi-space which permittivity is $\varepsilon=3.5$ (red dashed line) and on  surface of PEC-backed dielectric substrate  for two different thickness 25~$\mu$m (blue solid line) and 10~$\mu$m (green dot-dashed line),  $\mu_c=0.5$~eV.} \label{fig:SPPre}
\end{figure}

\begin{figure} 
\centerline{\includegraphics[width=.95\columnwidth]{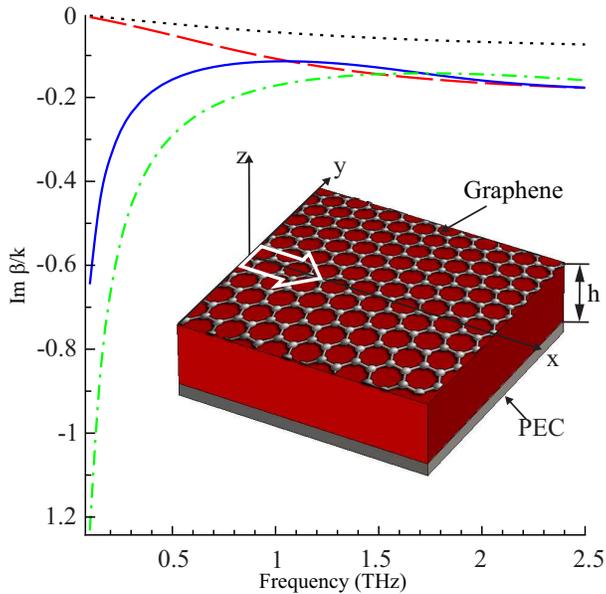}}
\caption{Frequency dependences of imaginary parts of propagation constants of SPPs in  same cases as  in Fig.~\ref{fig:SPPre}.} \label{fig:SPPim}
\end{figure}

In  case of the space $h$ between the graphene sheet and PEC is infinitely large, the dispersion equation for the SPP propagating along graphene placed on a dielectric semi-space follows from (\ref{eqMS}) and can be written as
\begin{equation}
\gamma_1 \varepsilon + \gamma_2 (1+\sigma \eta \frac{\gamma_1}{k})=0. \label{eqDS}
\end{equation}
The well known expression for the propagation constant for the SPP supported by a free standing graphene layer \cite{hanson-2008-JApplPhys} follows from  equation (\ref{eqDS}):
\begin{equation}
\beta=k\sqrt{1-\frac{4}{\sigma^2 \eta^2}}\label{eqFS}
\end{equation}
where we put $\varepsilon=1$. The frequency dependences of the corresponding SPP propagation constants are shown in  Fig.~\ref{fig:SPPre} and Fig.~\ref{fig:SPPim}.

\begin{figure} 
\centerline{\includegraphics[width=.95\columnwidth]{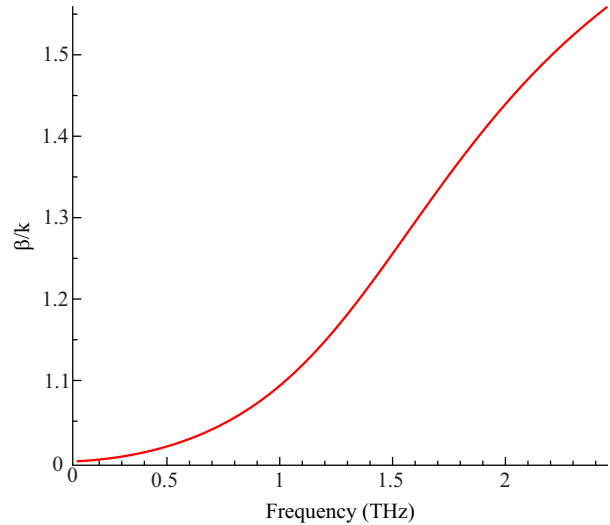}}
\caption{Propagation constant of fundamental TM mode of PEC-backed dielectric layer ($\varepsilon=3.5$) of 25~$\mu$m thickness versus frequency.} \label{fig:TMbs}
\end{figure}

Lastly, in case of the conductivity $\sigma$  equals to zero, i.~e. the interface $z=0$ is free from graphene, the dispersion equation for the TM eigenwaves follows from (\ref{eqMS}) and can be written as 
\begin{equation}
\sqrt{k^2\varepsilon-\beta^2}\tan{(h\sqrt{k^2\varepsilon-\beta^2})} =\varepsilon\sqrt{\beta^2-k^2}. \label{eqSB}
\end{equation}
If PEC-backed dielectric layer is non dissipative, at least one real solution of  equation (\ref{eqSB}) exists corresponding the fundamental lowest frequency TM mode which has no cut-off frequency. The frequency dependence of the mode propagation constant is shown in Fig.~\ref{fig:TMbs}.

\section{Analysis of resonant properties of graphene reflectarray}

In this Section, we consider a case of  array excitation  by normally incident $y$-polarized plane wave (see Fig.~\ref{fig:Structure}a). In the  conclusions we shall give short comments related to $x$-polarization of the wave. 
From the dispersion dependences presented in Figures~\ref{fig:SPPre},~\ref{fig:SPPim} one can see that  retardation of the eigenwave of non-patterned graphene (on-dielectric PEC-backed structure) is  larger for the dielectric layer with  smaller thickness. But reducing  the structure thickness results in increasing of the wave damping (see Fig.~\ref{fig:SPPim}). Thus we have chosen the dielectric substrate thickness $h=25$~$\mu$m which is typical for THz devices.  

\subsection{Features resulted from resonant dimensions of graphene strips}

The reflectivity of structure manifests in a resonant deep resulted from  excitation of SPPs with wavelength which is approximately equal to the length of graphene strip inside a unit cell. 
As a consequence of the structure symmetry, the incident $y$-polarized wave does not excite currents at the edge points of the strip where the electric field is orthogonal to the strip. To estimate the value of resonant frequency, we compare the length of the line drowned between points $A$ and $B$ (they are placed on the boundaries of unit cell, see Fig.~\ref{fig:Structure}a) with zero values of surface current density and the wavelength of SPP by data of solution of dispersion equation (\ref{eqMS}) (see Fig.~\ref{fig:SPPre}). Approximate resonant condition is reduced to equality of the distance between points $A$ and $B$ to a half of the SPP wavelength. This estimation leads to the resonant frequency 0.563~THz.

\begin{figure} 
\centerline{\includegraphics[width=1.0\columnwidth]{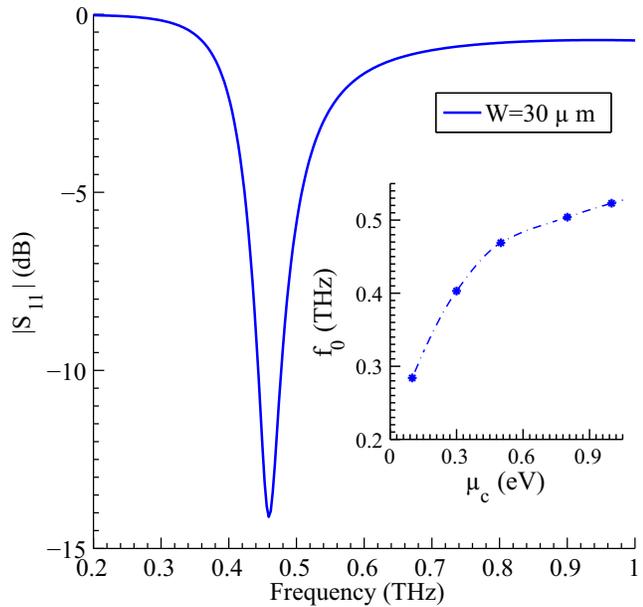}}
\caption{Frequency dependence of reflection coefficient for $h=25$~$\mu$m, $w=30$~$\mu$m, $\mu_c=0.5$~eV, normally incident $y$-polarised wave. Insert: dependence of resonant frequency  on chemical potential $\mu_c$.} \label{fig:spectr1_Y}
\end{figure}

Frequency dependence of reflection coefficient obtained by full waves numerical simulations is presented in  Fig.~\ref{fig:spectr1_Y}. If one  takes into account a complexity of the shape of the array elements and the essential decay factor of SPPs, one can see that our analytical estimation of the resonant frequency and the related numerical data are in a good agreement. Thus, this lowest frequency resonance of the reflectarray depends on the strip dimensions and it is similar to resonances of classical microstrip resonators and antennas in microwave region \cite{pozar-1992-ProcIEEE} and also resembles resonances of longitudinal currents in the fish-scale array with narrow PEC strips \cite{prosvirnin-2006-APL}. In our case,  the  resonant surface current distribution is shown in Fig.~\ref{fig:Structure}b. 

The propagation constant of SPPs depends on graphene chemical potential, therefore the resonant frequency of the structure can be changed by an external bias voltage. In the insert of Fig.~\ref{fig:spectr1_Y} we present the  dependence of the resonant frequency on the  chemical potential $\mu_c$.
\subsection{Resonances resulted from  excitation of eigenmodes}
In  periodic structures, the scattered field of normally incident plane wave has discrete space spectrum and can be presented as a superposition of the principal waves propagating along $z$-axis and partial waves (propagating or evanescent) both inside the substrate  and above the graphene in the near free space. The lowest partial waves propagate in  plane $xOz$ (one pare of waves) and $yOz$ (another pare). They have identical absolute value of $x$- and $y$-components of the wave vectors $2\pi/l$.

To observe resonances of reflection based on energy store in electromagnetic field oscillation resulted from  excitation of opposite propagating eigenwaves, it is necessary to fulfill the following requirements. In the plane $xOy$ the propagation constants of both  eigenwave of the structure $\beta$ and the partial wave of the scattering field must be identical:     
\begin{equation}
\beta(f)=2\pi/l.\label{synchronism}
\end{equation}
It is a synchronism condition. 

Next, to have a high quality factor of resonance, it is necessary to exclude  energy loss via in plane  propagating eigenwaves. Thus the amplitudes of excited eigenwaves along some direction and in opposite one must be identical. Let us consider firstly the excitation of eigenmodes propagating in $x$-direction. The amplitudes of excited eigenwaves along $Ox$ axis and in the opposite direction must be equal. This requirement is fulfilled in  case of a fish-scale structure because the plane $yOz$ is its symmetry plane. In  case of eigenwaves propagated along $y$-direction, the condition is more complex because the plane $xOz$ is a glide symmetry plane of the array. However, it is sufficient condition to exclude  a direct or reversal power flow. 

\begin{figure} 
\centerline{\includegraphics[width=1.0\columnwidth]{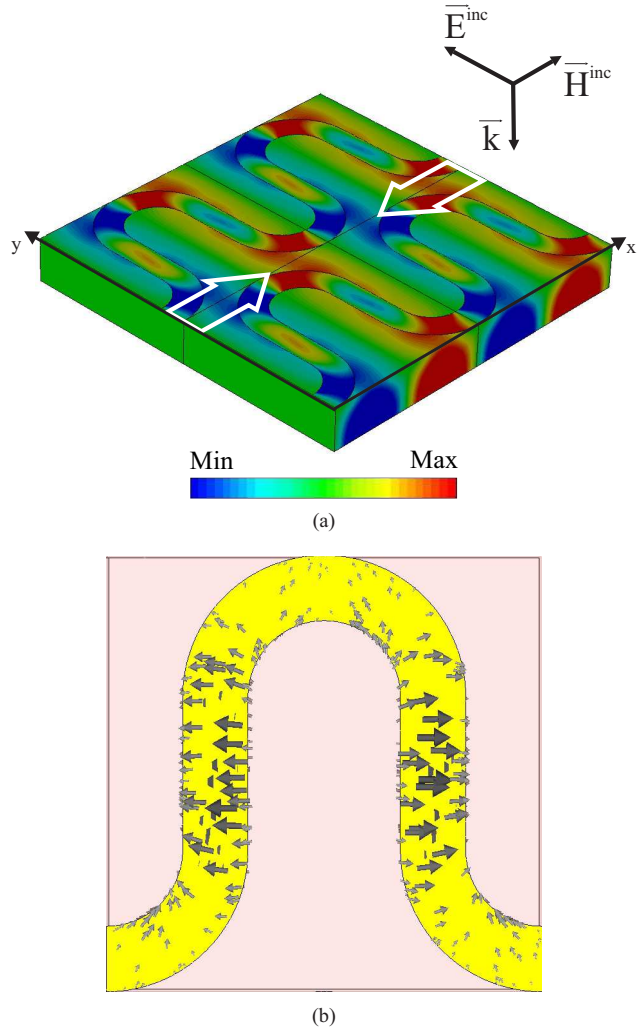}} 
\caption{a) $H_y$-component of resonant oscillation ($f$=1.887~THz), eigenwaves propagate in $x$-direction,  excitation by $y$-polarized normally incident wave, $h=25$~$\mu$m, $w=15$~$\mu$m, $\mu_c=0.5$~eV. b) Surface current distribuition in this case} \label{fig:H_y}
\end{figure}

Thirdly, the phases of the opposite propagating eigenwaves must correspond to constructive interference of their fields inside the substrate with the scattered field of the incident wave. This  condition is satisfied also due to the shape of graphene strip in the unit cell, as it can be seen easily. For instance, let us consider the case of eigenwaves propagated along $x$-axis. The longitudinal current has zero values in the points  dividing strip inside a unit cell in two parts (they are the points $A$ and $B$ shown in the Fig.~\ref{fig:Structure}a). These two parts are $S$-shaped planar chiral particle and mirror to $S$-shaped one with opposite handedness. Thus, on the surface of these particles the longitudinal currents induced by incident $y$-polarized field excite magnetic fields having opposite directed  $H_y$. These fields are shifted in $l/2$ along $x$-axis that is the same as right- and left-handed $S$-shaped particles inside unit cell (see Fig.~\ref{fig:H_y}a).

It is also worth to  mention  that the main field components of the excited resonant oscillation which are $H_y$, $E_x$ and $E_z$, are absent in the incident plane wave. As a consequence, coupling of the oscillation and the incident wave is small and the radiation loss is also small. Thus we deal with the trapped mode regime of the open planar periodic structure. 
For the sake of brevity we will not now consider in detail the case of resonant excitation of eigen waves propagated in $y$-direction. 
Thus the fish-scale patterned structure is a perfect object to observe resonant features resulted from excitation of eigenmodes of planar photonic crystal.

\begin{figure} 
\centerline{\includegraphics[width=1.0\columnwidth]{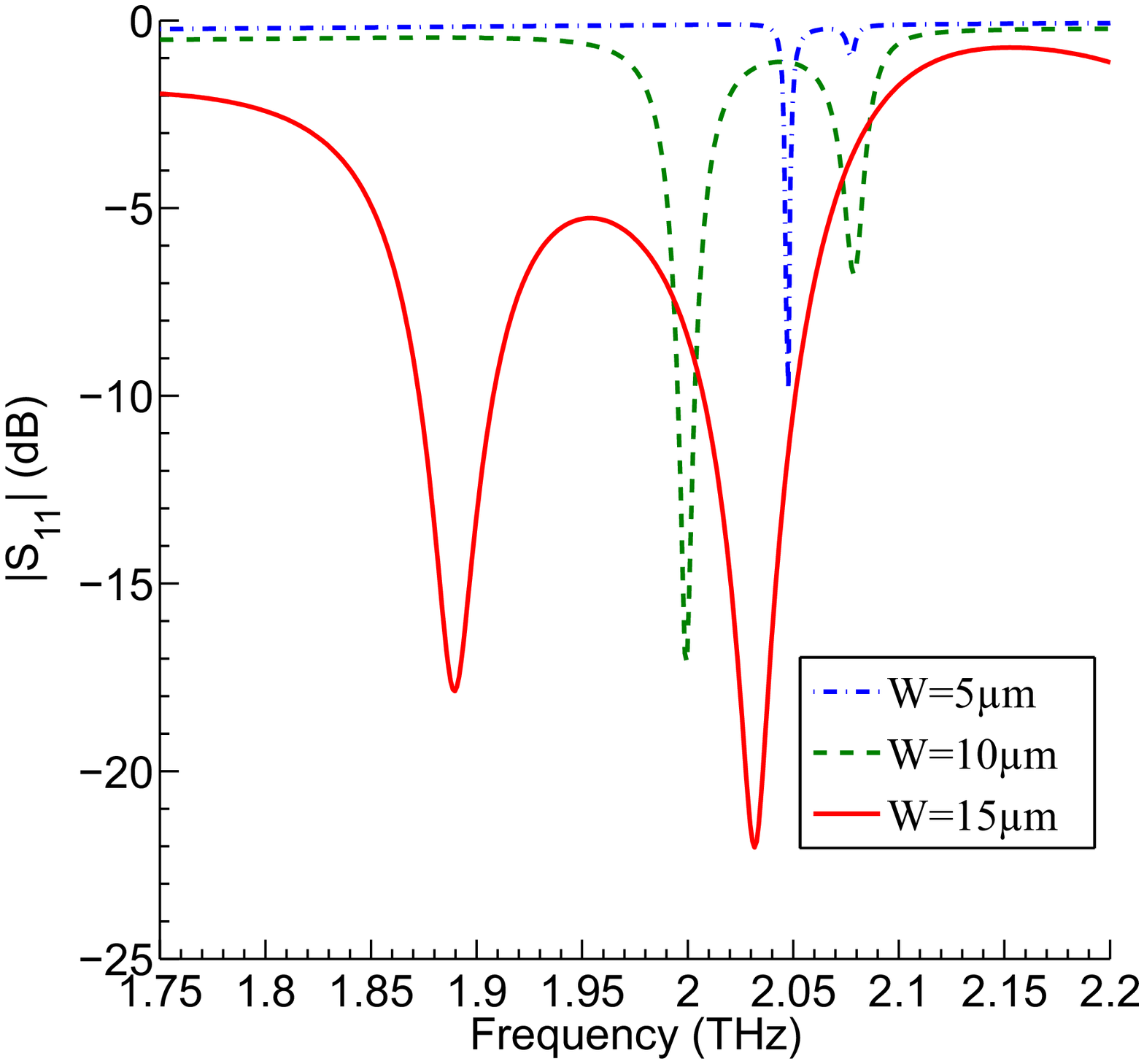}}
\caption{Frequency dependence of reflection coefficient: $h=25$~$\mu$m, $\mu_c=0.5$~eV, normally incident $y$-polarized wave.} \label{fig:spectr2_Y}
\end{figure}

\begin{figure} 
\centerline{\includegraphics[width=1.0\columnwidth]{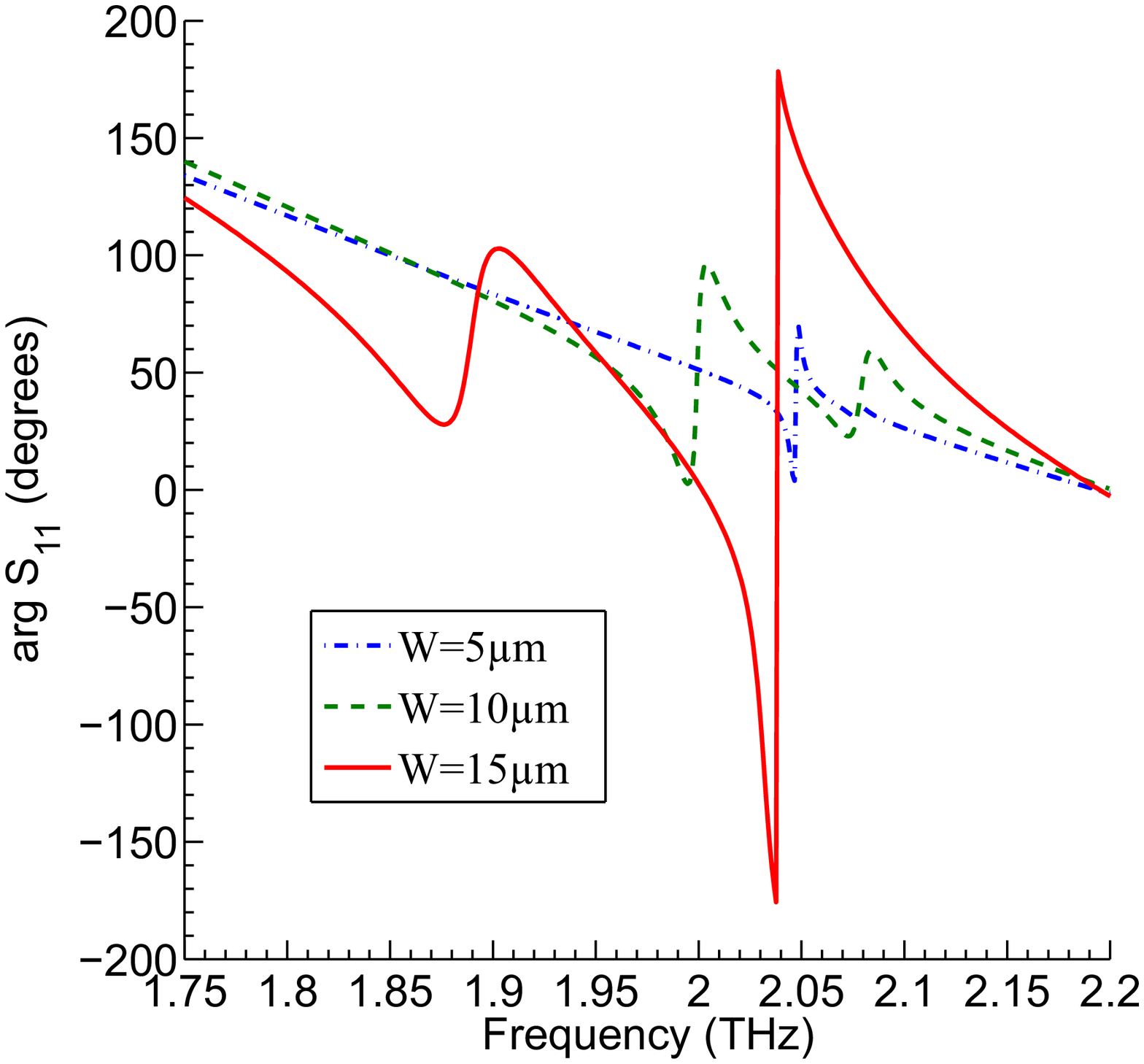}}
\caption{Frequency dependence of phase of reflection coefficient: $h=25$~$\mu$m, $\mu_c=0.5$~eV, normally incident $y$-polarized wave.} \label{fig:phase_Y}
\end{figure}

The spectrum of reflection has been obtained numerically and it is shown in Figures~\ref{fig:spectr2_Y}, \ref{fig:phase_Y}.
\begin{figure} 
\centerline{\includegraphics[width=1.0\columnwidth]{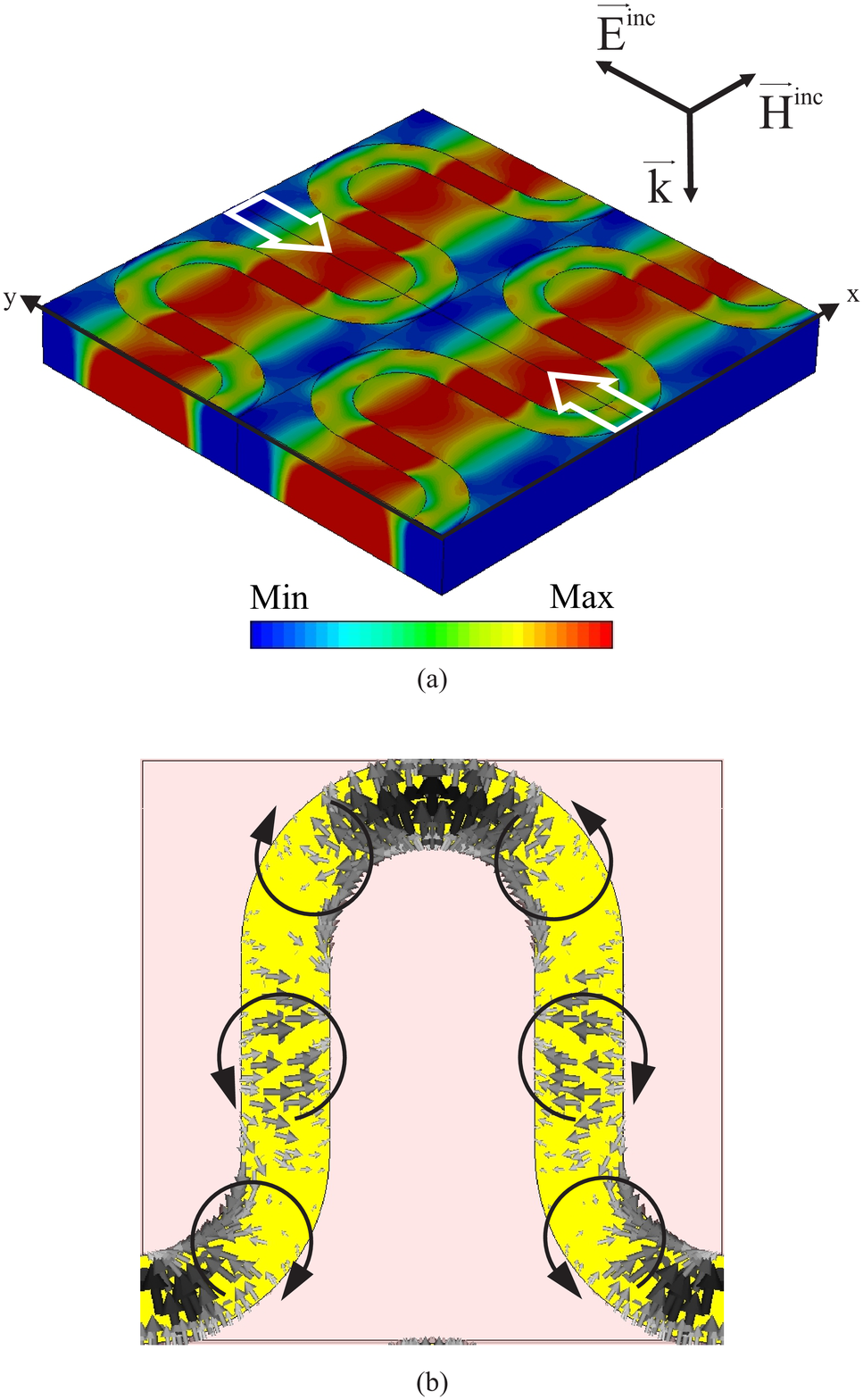}} 
\caption{$H_x$-component of resonant oscillation, $f$=2.031~THz, eigenwaves propagate in $y$-direction, excitation by $y$-polarized normally incident wave,  $h=25$~$\mu$m, $w=15$~$\mu$m, $\mu_c=0.5$~eV. b) Surface current distribuition in this case} \label{fig:H_x}
\end{figure}
As one can see, the resonances of these kind are excited by $y$-polarized wave as a doublet for every  of the chosen strip width. The resonant frequencies and the difference between them are collected in  Table~1. The shown data reveal that the narrow strips have close resonant frequencies in doublet because the features of the array with narrow strips are closer to the properties of PEC-backed dielectric as a waveguide. 

The lower resonant frequency of each doublet corresponds to the excitated eigenwaves propagating along $x$-direction and the higher frequency corresponds to propagating ones along $y$-axis. As the main component of plasmon-polariton magnetic field at lower resonant frequency is orthogonal to the same component of the incident wave, the quality factor in this case is higher than the one related to the second resonance where these components are parallel resulting in a bigger radiation loss. 

The transverse to strip surface current components prevail at the excitation of this kind of resonances (see Figures~\ref{fig:H_y}b and \ref{fig:H_x}b ) in contrast to the longitudinal currents at the resonances considered above in section~4.A. Besides,  in the second resonance of doublet, because the structure has glide symmetry plane being orthogonal to propagation direction of the eigenwaves, the related standing wave manifests a complex field distribution inside the unite cell. It is resulted in appearance of regions on graphene strips with specific rotating current distributions which are shown in Fig.~\ref{fig:H_x}b.  

\begin{table} 
  \caption{Resonant frequencies of doublet for different $w$, $\varepsilon=3.5$, $h$=25~$\mu$m, $\mu_c=0.5$~eV}
  \begin{center}
    \begin{tabular}{cccc}
    \hline
    $w$~[$\mu$m] & $f_1$ [THz] & $f_2$ [THz] & $\Delta f=f_2-f_1$ [THz] \\
    \hline
    5 & 2.049 & 2.076 & 0.0274 \\
    10 & 1.986 & 2.070 & 0.084 \\
    15 & 1.887 & 2.031 & 0.144 \\
    \hline
    \end{tabular}
  \end{center}
\end{table}

The quality factor of resonance depends on dissipative properties of the materials of the structure and on diffraction loss. For the sake of simplicity, we assumed that the dielectric substrate is nondissipative.

Features of eigenwave of the considered structure differs in a small way from those of eigenwave of PEC-backed substrate in the case of narrow graphene strips. The resonant frequencies  are far from those of the above studied dimensional resonances of the strips. Thus, the frequency of this photonic crystal kind of resonances can be estimated by using in the equation (\ref{synchronism}) instead of rigorous value $\beta$, an approximated one related to eigenwave of backgrounded substrate and presented in Fig.~   \ref{fig:TMbs}. The solution of approximate equation (\ref{synchronism}) provides  estimation of resonant frequency as  $f=2.057$~THz which is in  good correspondence with resonant frequencies following from numerical simulations (see Table~1). The frequency discrepancy is smaller when the strip width is narrower because  perturbation of the eigenwaves is smaller.  

In Table~2 we present data characterizing  dependence of the resonant frequencies of the doublet and absorption ($A=1-|S_{11}|^2$) on the  chemical potential $\mu_c$. One can see a smaller variation of resonant frequency in comparison with the discussed above dimensional resonance tuning. It is evident consequence of quite different physical nature of these kinds of resonances. However, by changing chemical potential due to variation of bias voltage, one can modify the level of absorption of the structure. It is important for application in  devices for modulaton of  signals \cite{2012-graphene-modulator}.

\begin{table} 
  \caption{Resonant frequencies and absorption  of doublet for different $\mu_c$, $\varepsilon=3.5$, $h=25$~$\mu$m, $w=15$~$\mu$m}
  \begin{center}
    \begin{tabular}{ccccc}
    \hline
    $\mu_c$~[eV] & $f_1$ [THz] & $f_2$ [THz] & $A_1$ [\%]  & $A_2$ [\%]  \\
    \hline
    0.3 & 1.722 & 2.142 & 99.6 & 70.8\\
    0.5 & 1.887 & 2.031 & 98.2 & 99.3\\
    0.8 & 1.923 & 2.118 & 98.3 & 99.3\\
    1.0 & 1.980 & 2.166 & 65.6 & 84.9\\
    \hline
    \end{tabular}
  \end{center}
\end{table}
\section{Conclusions}
We reported features of novel tunable graphene fish-scale reflectarray. Uninterapted strip elements of array are convenient to apply control bias voltage. 
The structure manifests two kinds of resonances which differ by physical nature. Analysis reveals that they are excited by incident wave in cases of polarizations both along the strips of array and transverse to them.  The first one exists because inside of periodic cells the strips of structure have resonant lengths for induced currents.  The reflectivity of structure manifests in a resonant deep (peak of absorption) resulted from excitation of SPPs with wavelength which is approximately equal to the length of graphene strip inside a unit cell in the case of excitation by incident wave polarized transversally to array strips. 

The second kind of resonances of fish-scale array placed on the metal-backed substrate is resulted from excitation of TM eigenwaves of the structure which have identical amplitudes but propagate in opposite directions. An energy is stored in the standing wave along the whole structure being a plane photonic crystal. These resonances are not dependent on whether the strips have resonant sizes within the unit cell or not.

\section*{Acknowledgments}
The authors would like to acknowledge  financial support of the  Brazilian agency CNPq.

\section*{References}
\bibliographystyle{unsrt}
\bibliography{GrapheneFSreferences}
\end{document}